
\vskip 3.cm
\centerline {\bf NUCLEAR CHARGE DENSITY DISTRIBUTIONS}
\centerline {\bf FROM ELASTIC ELECTRON SCATTERING DATA}
\vskip 2.cm
\centerline {R.Anni, G.Co' and P.Pellegrino}
\vskip .4cm
\centerline {Dipartimento di Fisica, Universit\`a di Lecce}
\centerline {and}
\centerline {INFN, Sezione di Lecce}
\centerline {I-73100 Lecce, Italy}
\vskip 1cm

{\bf Abstract.} \quad
The model independent procedure of extracting charge density
distributions from elastic electron scattering data is
investigated.  The charge density
distributions are expanded on an orthonormal basis and
the parameters of the expansions are fixed by the comparison with
the experimental data.
Two bases with different analytical properties (Fourier-Bessel and
Hermite) are used. This allows us to disentangle the
uncertainties coming from the choice of the expansion basis from
those intrinsic to the extraction procedure.
We design
a set of tests to select the number of
the expansion coefficients adequate for a proper description of the
data. The procedure is applied to elastic data measured on $^{12}$C,
$^{40}$Ca  and $^{208}$Pb nuclei.

\vfill\eject

{\bf 1. Introduction}
\vskip 0.5 cm

The advantages of using electrons in the
investigation of the nuclear structure are mainly related to
the fact that the electron-nucleus interaction is
relatively weak.
For this reason
multiple scattering effects are usually
neglected and the
scattering process is described in
terms of perturbation theory.

Since the reaction mechanism in perturbation theory is
well under control the connection between the cross section and
quantities such as charge
distributions, transition densities, response functions
etc., is well understood.
This is the reason
why the comparison between
theory and experiment is often done
not at the cross section level,
but using these quantities.

The techniques used to extract these quantities
from the cross section data are, in principle,
grouped under two basic approaches.
A first approach consists in the solution of the inverse
scattering problem, or in other words, in obtaining the
potential from the cross section data.
In the second approach one solves the direct
scattering problem making a guess about the form of the
potential and calculating the cross section.
The parameters fixing the form of the potential are
changed until the experimental cross sections are
reproduced.

Neither of these procedures is trouble free.

The inverse scattering approach applied to a set of
experimental data, which are measured on a
discrete grid and have error bars, does not ensure a unique
solution because of numerical instabilities
[1]. This makes the use of this
approach extremely impractical and
for this reason it has never been applied
in this context.

On the other hand, the direct scattering approach is limited by the need
of guessing an analytical expression of the potential, or of the quantity
generating it.

In the past good descriptions of the experimental data were obtained
with analytical expressions of the density depending from few
parameters (e.g the Fermi distribution). In these cases the parameters
are related  to the characteristics of the shape of
the density, for example the half-density radius, the diffuseness, the
central depression. The development of the accelerator and detector
technology has produced data of better quality and covering an increasing
range of momentum transfer. The simple, and intuitive, analytical expressions
mentioned above are not able to provide satisfactory descriptions of these
new high quality data. This has been achieved, within the direct
scattering approach, with the introduction of the so-called model
independent methods.

These methods are based upon the assumption that
the function describing the density can be expanded on a complete set of
orthonormal functions. The coefficients of the expansion are changed in order
to reproduce the  experimental cross section. Since the basic assumption
is very general (it is hard to believe that functions describing density
distributions or quantities related to them like potentials have such odd
analytical properties which prohibit them to be expanded on a basis)
from the mathematical point of view the success of these methods is
guaranteed.
The mathematical certainty of the success of these methods should however
pay the price of two shortcomings: first the parameters of the expansion
are not expected to have a particular physical meaning, and second,
in practical applications the expansion must be truncated to a finite
number of terms and this generates uncertainties in the
determination of the distribution.

The uncertainties  related to the
extraction procedure should be added to those related to
the experimental errors.
Our claim is that, given the high accuracy
of the modern electron scattering experiments, these theoretical
uncertainties could be of the same order of magnitude, or
even bigger, than those coming from the experimental errors.

In this paper we present the results of our study on
the determination of the charge density distributions of
doubly-magic nuclei from elastic scattering processes.
Specifically, we have been interested in investigating
the source and the magnitude of the theoretical
uncertainties linked
to the model independent extraction techniques.

To obtain information independent from the
choice of the expansion basis, we have used
two  orthonormal bases with quite
different analytical properties.
These are the Fourier-Bessel (FB) basis, widely used
in the literature to
analyze both elastic and inelastic experimental data
[2, 3, 4], and a
basis of Hermite functions which, as far as we know,
we are using for
the first time in this context.

In the next section of the paper we describe the
properties of the two bases and the
mechanisms which allow them to reproduce the cross section.
For simplicity this study is done in
Plane Wave Born Approximation (PWBA).

In section 3 we use the two bases to reproduce pseudo
data generated by model charge density distributions
as well as really measured cross sections. This analysis
is done solving the Dirac equation for an electron
scattered by a Coulomb potential in the ultra-relativistic
limit. We present a comparison between the performances of
the two bases in reproducing the data and we discuss the
common features and limitations. We discuss the
propagation of the experimental errors from the cross
section to the density.
In this section we analyze the uncertainties related to the
truncation of the expansion basis and
we define a procedure to fix the number of expansion
coefficients.

Finally, in the last section we summarize the work and we
draw our conclusions.

\vskip 1 cm
{\bf 2. Model independent analysis.}
\vskip 0.5 cm

Our study of the determination of the charge
distributions from elastic scattering processes is
done solving directly the scattering problem. The
potential is generated by a trial charge distribution
whose parameters are chosen to reproduce
the experimental cross section.

In the model independent procedure
the trial density distribution is expanded on
an orthonormal basis $P_n(r)$
$$
\rho(r) = \sum_{n= 1}^{\infty} A_nP_n(r).
\eqno(2.1)
$$
and the
expansion coefficients $A_n$ are fixed to reproduce the cross
section.

This is the straightforward application of the model
independent method. In reality, this method is more general, and can
be applied to any function which is determining the density in a
unique way. For example, in the generalized Helm model of ref. [5]
the density is generated as a convolution of two terms, one of these
(the so-called Helm density) is expanded on a basis of Legendre
polynomials.

In Plane Wave Born Approximation (PWBA) the
link between the charge density distribution and the cross
section is straightforward.
In this work we shall limit our
considerations to  even-even nuclei. In this case,
defining  the form factor as
$$
 F(q) = 4 \pi \int_{0}^{\infty} \, dr r^2 j_0(qr) \rho(r),
\eqno(2.2)
$$
where $j_0$ is the spherical Bessel function of zero order
and $q$ the absolute value of  the
three momentum transfer, the cross section can be written
as:
 $$
{ d \sigma (q) \over d \Omega }= \sigma_M f_{rec}
| F(q) |^2.
\eqno(2.3)
$$

In eq. 2.3 we have indicated with $\sigma_M$ the Mott cross
section and with $f_{rec}$ the recoil term:
$$
f_{rec}= (1+ {2 \epsilon \over M_A} sin^2 ,
{\theta \over 2} )^{-1}
$$
where
$\epsilon$ is the electron energy, $M_A$ is the
mass of the nucleus and $\theta$ is the scattering angle.

In PWBA the inverse scattering problem can be easily
solved. It is possible to extract the form factor from the
cross section and then, with an inversion of the Fourier
transform, to get the charge density distribution
$$
\rho(r) = {1 \over 2 \pi^2}
\int_0^\infty j_0(qr) F(q) q^2 \,\, dq .
\eqno(2.4)
$$

The intrinsic limitations of the direct
and inverse approaches in the determination of the charge
density distribution are well shown by the above
equations.

Eq. 2.1 shows that the direct approach requires an
infinite sum of the expansion terms. From
the pragmatical point of view this sum has to be truncated
and this produces an error in the determination of the
charge density distribution.

On the other hand, it is shown by eq. 2.4 that,
in the solution of the inverse scattering problem, a
precise determination of the charge density distribution
is connected to the knowledge of the cross
section up to infinite momentum transfer.
The measurements are clearly performed
in a limited range of $q$ and this produces an
uncertainty in the determination of the charge density
distribution.

The relation between these intrinsic limitations of the two
procedures can be better seen by specifying the expansion
basis of the density.

In our study we made the hypothesis that the charge
distribution is a continous and even function of the
distance $r$ from the center of the nucleus.
This means that an analytical continuation
of the density in the unphysical region of the negative distances
would provide $\rho(-r)=$ $\rho(r)$.

\vskip 1 cm

{\sl 2.1 The Fourier Bessel expansion}
\vskip 0.5 cm

The Fourier Bessel (FB) basis has been widely used to analyze both
elastic and inelastic data [4], and here we simply sketch the
basic ideas underlying the expansion procedure referring to ref.
[2,3] for further details.

The basic idea of the FB expansion is related to the
possibility of expanding a single valued
function defined in a finite domain.
For this purpose one uses the
orthogonality relation between spherical Bessel functions
in a finite domain:
$$
\int_0^{R_c}\,dr r^2 j_l(q_nr )
j_l(q_m r ) = { R_c^3 \over 2}
[j_{l+1}(q_n R_c)]^2 \delta_{n, m} ,
\eqno(2.5)
$$
where the $q_n$ are defined such as
$$
j_l(q_n  R_c) =0.
\eqno(2.6)
$$

In our case the use of the FB basis implies that the
charge density $\rho(r)$ should be zero for values
of $r$ larger than $R_c$:
$$
\rho(r)= \cases {\sum_{n=1}^{\infty} a_n
j_0(q_nr) & for $r \leq R_c$ \cr
0            & for $r > R_c$ ,\cr}
\eqno(2.7)
$$
where, from eq. 2.6,
$$
q_n={ n\pi \over R_c}. \eqno(2.8a)
$$

In principle it is possible to obtain the $a_n$
coefficients measuring directly the cross section at the
$q_n$ momentum transfer.
If the form factor 2.2 is known in $q_n$,
the coefficients $a_n$ can be obtained
inserting of the relations 2.5 and 2.7 in the definition 2.2 of the
form factor:
$$
a_n={ F(q_n) \over 2 \pi R_c^3 [j_1(q_nR_c)]^2 } .
\eqno(2.9)
$$

In general the cross section is
measured at $q$ values different from $q_n$.
Using the expansion 2.7 of the charge density we find for the
form factor the relation:
$$
F(q)= {4 \pi R_c^2 \over q} \sum_{n} a_{n}
{ (-1)^n\over (qR_c)^2-(n\pi)^2 }
sin(qR_c).
\eqno(2.10)
$$

Before applying the FB expansion of the density to realistic
cases we illustrate in a simple example the mechanism which allows
this  expansion to reproduce the form factors.
In order to keep the calculations as simple as possible
we have performed
this study in PWBA. The form factor that the FB
techniques should reproduce has been generated by a
density of analytical form. Specifically we have used
densities given by symmetrized Fermi distributions [6]
$$
\rho(r)= \alpha {cosh(R/d) \over  cosh(R/d) +
cosh(r/d)  }\,\, .
\eqno(2.11)
$$
For this density an analytical expression of
the form factor can be easily obtained [7]:
$$
F(q)= - { 4 \pi^2 \alpha d \over q} { cosh(R/d) \over
sinh(R/d) } \Biggl[ {R cos(qR) \over sinh(\pi qd) } -
   {\pi d sin(qR) cosh(\pi qd) \over sinh^2(\pi q d) }
\Biggr].
\eqno(2.12)
$$

The above expression is composed of
oscillating terms damped by exponentials. This is better seen
taking the limit for $qd \gg 1$. We obtain:
$$
F(q) \simeq - { 4 \pi^2 \alpha d \over q}
[ R cos(qR)  - \pi d sin(qR) ] e^{-\pi q d},
\eqno(2.13)
$$
where the oscillating and damping terms are evident.

Only two of the above parameters $\alpha$, $R$ and $d$  are
independent since the normalization condition
$$
4 \pi \int dr \, r^2 \, \rho(r) = Z
\eqno(2.14)
$$
relates them.

In the panels a) and c) of
fig.1, together with the form factor given by eq.(2.12)
(thick full lines), we present the  results obtained using respectively $4$
and $6$ FB coefficients calculated from eq. (2.9).
The thin
lines represent the results obtained with 4 and 6 FB expansion coefficients
and the contribution of the last coefficient of the expansion is shown
by the dashed lines.

The contribution to the form factor of
the n-th expansion term  is localized around the $q_n$ value of eq.
(2.8a).
Let us consider for example the sixth term. Its
major contribution is peaked around $q=2.5$ $fm^{-1}$, and it
oscillates on the full momentum transfer range. For $q<2.5$
$fm^{-1}$ the contribution of this term is orders of magnitude
smaller than the full form factor, therefore irrelevant for the fit.
The presence of this term is however necessary to reproduce the form
factor around $q=2.5$ $fm^{-1}$. For larger values of $q$ the sixth
term presents an oscillating behavior completely different from
that of the
true form factor, and the fit above this value is
no longer reasonable.

This analysis confirms the naive idea that
the higher expansion coefficients of the charge density
distribution influence only the high $q$ part of the form
factor. There is therefore a link between the experimental
limitation of the data at a maximum value of the momentum
transfer and the truncation of the FB expansion.
The number of expansion coefficients determined by
experiments performed up to $q_{max}$ can be obtained
inverting eq. 2.8:

$$
n_{max}= { q_{max} R_c \over \pi}.
\eqno(2.8b)
$$

The use of the FB expansion implies a
strong hypothesis on the shape of the density forcing it
to be zero for $r>R_c$. This is producing a
discontinuity in the analytical shape of the density with
consequences on the form factor.

In fig.2 we compare the form factors calculated for
the cases of  $^{12}C$, $^{40}Ca$  and $^{208}Pb$
nuclei using the exact expression (2.12) with those obtained truncating the
density (2.11) at different values of $R_c$. The calculations have
been done with the densities of eq. (2.11) with  $R=1.1$ $A^{1 \over
3}$ $fm$ and $d=0.626$ $fm$.

The thick lines are representing the squares of the form factors
calculated with eq. (2.12). The thin lines show the square of the
form factor calculated with the truncated density distribution
($F_c(q)$). An estimate of effect of the truncation on the
form factor is given by the expression:

$$\eqalignno{\Delta F(q) \equiv  F(q) - & F_c(q)
=  {4 \pi \over q} \int_{R_c}^\infty dr r sin(qr) \rho(r) \cr
\simeq {8 \pi \over q} d^2 \alpha cosh( {R \over d})  & \Bigl\{
 { e^{-{R_c \over d} } \over (1+(qd)^2) }
\Bigl[ (qd) cos(qR_c)({R_c \over d}(1 +(qd)^2) + 2) \cr
+  sin(qR_c) & ({R_c \over d}(1+(qd)^2)+1-(qd)^2
\Bigr] \Bigl\} , &(2.15)}
$$
which results in all the cases here considered in excellent
agreement with the exact result obtained with a numerical
integration.

The difference between the two form factors is not spread over the full
range of the momentum transfer but it shows up at a certain value of
$q$ after which the true form factor
decreases while the form factor obtained with the truncated density
oscillates around a roughly constant value.

Eq. 2.15 shows that $\Delta F(q)$
becomes smaller the larger is the
value of $R_c$.

$R_c$ can be fixed by the condition that
for a given $q$  the ratio $\Delta F(q) / F(q)$ (eqs. 2.15 and 2.12)
should be smaller than a fixed value.
For example, the $R_c$ radii used in
fig. 2  are $8$ $fm$ for both $^{12}C$ and $^{40}Ca$,  and $11$
$fm$ for $^{208}Pb$.

The truncation of the density generates an
unphysical contribution which,
for large values of $q$, dominates on the true value
of the form factor.
Using the FB expansion one should take care to choose the
$R_c$ parameter such that
the unphysical contribution remains small in
the range of $q$ values investigated.

\vskip 1cm
{\sl 2.2 The Hermite expansion}
\vskip 0.5 cm

In order to disentangle the uncertainties related to the
choice of the expansion basis from those intrinsic of the extraction
method we have worked with
two bases having different analytical properties.
In addition to the FB expansion presented above,
we have used a basis of Hermite functions [8]:
$$
u_n(x) = ( \sqrt \pi 2^nn! )^{ -{1 \over 2} }
e^{- {x^2 \over 2} }
H_n(x) ,
\eqno(2.16)
$$
where we have indicated with $H_n$ the Hermite
polynomial of  order $n$.

The orthogonality relation of the $h_n$ functions is
given by:
$$
\int_{-\infty}^{\infty}
u_n(x) u_m(x) \, dx = \delta_{n,m}.
\eqno(2.17)
$$

The charge density distribution is expanded as:
$$
\rho(r)= \sum_{n=0 \atop n \, even}^\infty
c_n u_n( {r \over \beta}
) ,
\eqno(2.18)
$$
where $\beta$ is an arbitrary quantity, with the units of a
length, fixing the scale of the expansion basis.
The Hermite expansion removes the strong hypothesis done
by the FB expansion that the density should be zero for $r>R_c$.

The
expression of the form factor obtained inserting in eq.
(2.2) the Hermite expansion (2.18) is given by:
$$
F(q)= { \pi^{ 3/2} \over q} 2 \beta^2
\sum_{n=0 \atop n \, even}^\infty
\Bigl[ c_{n+2} \sqrt {n+2}
+ c_n\sqrt{n+1} \Bigr] (-1)^{n/ 2}
u_{n+1}(q\beta).
\eqno(2.19)
$$

The sum in eq. (2.18, 2.19) is restricted only
to the even values of $n$, since we have assumed that $\rho(r)$ is
an even function of $r$.

Like in the case of the FB expansion, we have studied the
mechanism which enables the Hermite basis to reproduce
the form factors.
In the panels b) and d) of fig.1 the thick lines are the squared
form factors obtained with the symmetric Fermi distribution,
eq.(2.11), the thin lines have been obtained summing respectively $4$ and $6$
terms of the Hermite expansion, and
the dashed lines represent the contribution of the last
term.

The comparison between the figures relative
to the FB calculation with the
analogous figure for the Hermite expansion shows a
remarkable difference
between the two bases.  The n-th term of the FB basis is oscillating
on the full range of $q$ and it presents a pronounced peak
in correspondence with the $q$
value where it gives its  main contribution to the full form
factor.
The n-th term of the Hermite basis is oscillating only up to
a maximum value of $q$, corresponding to the value where it gives
its main contribution to the full form factor, and, after, it
decreases without oscillations. Contrary to the FB case the
contribution of the n-th term does not show any pronounced peak.
It seems that, while in the FB expansion
there is a localization of the contribution of each expansion
term,
in the Hermite case the form factor is built up by
the calibrated sum of several terms.

Also in the case of the Hermite expansion the number of
expansion coefficients is related to the
maximum value of the experimental momentum transfer.
Fig. 1 shows that the addition
of new terms enables one to reproduce
the form factor on
a wider range of momentum transfer.
Finding a rule linking the number of coefficients
with the maximum value of the momentum transfer is not
straightforward as in the FB case where the coefficients are
directly related to $q_n$. In order to have an empirical
estimate of the maximum number of expansion coefficients
fixed by a given $q_{max}$ we have imposed that the
last inflection point of the form factor produced
by the n-th coefficient should lie at higher
momentum than $q_{max}$. With this assumption we
found the approximate relation:
$$
n \sim 0.5(q_{max}^2 \beta^2).
\eqno (2.20)
$$

\vskip 1cm

{\bf 3. Specific applications}
\vskip 0.5 cm

In the determination of a quantity, such as the
nuclear charge distribution, from a measured
quantity, the cross section, experimental and
theoretical uncertainties are both contributing.
In the specific case we are discussing,
the experimental uncertainties on the charge density are
related to the errors on the data and to the fact that the
cross section is measured on a limited range of momentum
transfer values.  The theoretical uncertainties are linked to
the use of a limited number of expansion coefficients which
produces also a dependence upon the choice of the expansion
basis.

The two bases described in the previous
section have been used
to investigate the theoretical
uncertainties of the charge distribution.

So far we have studied
the properties of the two bases
in the framework of the PWBA, but a realistic
description of the elastic electron scattering cross
section requires the full solution of the Dirac
equation. The Dirac equation for the elastic scattering of
electrons from a potential has been solved in the
ultrarelativistic limit, i.e. neglecting the terms
containing the electron mass, using standard procedures
[9].

The uncertainties of the fit procedures have been studied
working on
a set of pseudodata generated by known charge densities.
We have evaluated the ability of the expansion techniques to
reproduce the known densities and we have designed a procedure
to minimize the theoretical uncertainties.
In a second step we have applied this procedure
to analyze the experimental data on $^{12}C$,
$^{40}Ca$ and $^{208}Pb$ [10--13].

\vskip 1 cm
{\it 3.1 Application to pseudodata}
\vskip 0.5 cm

The set of pseudodata we have used to study the uncertainties of
the fit procedures
has been generated by shell model densities
produced by a Woods-Saxon potential:

$$ V_{WS}(r)={-V_0
\over 1+e^{(r-R)/a} }+ \left( { \hbar \over m_{\pi} c}
\right)^2 {1 \over r}  {d \over dr} \left( {-V_{ls} \over
1+e^{(r-R)/a} } \right) \; {\vec l} \cdot \vec \sigma +
V_{Coul}
\eqno(3.1)
$$
where we have indicated with $m_{\pi}$ the pion mass.
In the following we shall call these
charge distributions Woods-Saxon densities.

The values of the parameters used in our calculations are
given in Tab.1.

With the electrostatic potentials generated by these
charge distributions we have calculated the elastic cross
sections  at the same angles where the experimental points of
ref. [10--13] have been measured. We have
associated to each point the percentage error of the
analogous experimental point.  More information about the
pseudodata are given in Tab.2.

The fit procedure of the pseudodata consisted in modifying
the N expansion
coefficients of the charge distribution until
the minimum $\chi^2$ was found.

We should remark that not all the N coefficients are
independent. The normalization of the charge density, eq.
(2.14), imposes a condition on the parameters.
For the FB expansion, we have imposed the further condition
that the first derivative of the density should be zero at
$R_c$. In the following, if not explicitly mentioned, we
shall refer to the N expansion coefficients keeping in mind
that for the FB expansion only N-2 coefficients are
independent and for the Hermite expansion N-1. In the fitting
procedure only the  independent coefficients are changed.

We show in figs. 3 and 4
the sensitivity of the fit to the scale parameters $R_c$ and
$\beta$. In these figures the full lines represent the
Woods-Saxon density used to generate the pseudodata.

We have already discussed in sect. 2
the method used to fix $R_c$, and for $^{12}C$ with  data up to
$q_{max} \sim 3.0$ $fm^{-1}$ we found values of the cut-of radius
larger than $8$ fm.

The value of $\beta$ is chosen imposing two conditions
on the n-th term of the Hermite expansion: the
first one is that
the last inflection point in its contribution to the form factor is
beyond $q_{max}$, see sect. 2.2.
The second condition is asking
that the last inflection point of its contribution to
the charge distribution
should be far enough from the nuclear radius $R$ in a position $R_c$
where the density is negligible.

After imposing these two conditions we
found:  $\beta = \sqrt{R_c /q_{max}}$. For $R_c$ we choose the values
of cut off radius used in the FB case, therefore we obtained $\beta
\sim 1.5, 1.7$ $fm$.

We compare in fig. 3 two fits of the $^{12}C$
pseudodata done with a FB basis using
two values of $R_c$.
In this figure the Woods-Saxon density
used to generated the pseudodata is shown by the full lines.

In the upper panel of fig.3 we show the fit done with
$R_c=8.0$ $fm$. With 8 FB coefficients the charge distribution
is already well described as well as the cross section
pseudodata,
(we obtain a $\chi^2$ of about $1$ per degree of freedom).
Increasing the number of coefficients improves the
agreement with the true density. The situation remains stable
until we used 14 coefficients. In this case, while the
description of the cross section remains very good (
still we get $\chi^2$ $\sim 1$ per degree of freedom),
the density obtained shows unreasonable oscillations around the
true density.

The results of the lower panel of fig.3, where
we used $R_c=10$ fm, show that 8 FB coefficients are
not enough to obtain a reasonable description of the density
while the density obtained with 14 coefficients is still
adequate.

The general features of the fit procedure are moved to higher
order of coefficients when the value of $R_c$ is increased.
This is consistent with the empirical formula
$n_{max}=q_{max}R_c/ \pi$. A small value of $R_c$ allows one to
describe the data with a lower number of coefficients, on the
other hand, also the instabilities on the density show
up for a lower number of coefficients.

We reached the same conclusions also in the case of the
Hermite basis, shown in fig. 4.
In the upper panel of this figure
we present the results of the fits done
with $\beta=1.5$ fm, and we observe that a fit with 6
coefficient already reproduces very well the true
density.
In the lower panel we show the results of the fits done with
$\beta=1.7$ $fm$. In this case 6 coefficients are not enough to
reproduce reasonably well the density.

The comparison of fig. 3 and 4 seems to indicate that
the expansion on the Hermite basis is more stable than
the FB expansion. This means that, with respect to the FB
basis, it is possible to achieve a reasonable description of
the data, and of the true density, with a smaller number of
terms, and that the  unphysical oscillations
become evident only
for a larger number of terms. This is clear if one
compares the result of the fit with 16 terms of fig. 3 with the
analogous fit in fig. 4.

The two figures we have presented contain all the relevant
features of the model independent procedure for extracting charge
distributions from cross section data.

There is a minimum number of coefficients necessary
to get a reasonable fit to the data.
For a set of data, this minimum number depends on the
choice of the expansion basis and on the
value of the scale parameter.

There is also an upper limit of the number of coefficients
which allows one to produce reasonable densities
distributions.
Adding coefficients to the expansion maintains
constant the quality of the fit of the cross section but
it produces oscillations of the density.
This fact is easily understood looking at the mechanism
which allows the FB and the Hermite expansions to
reproduce the cross section (see fig. 1). High order
terms are effective at high values of the momentum
transfer. Therefore all the terms contributing in a
momentum region not covered by the data are undetermined.
They do not effect the fit of the cross section but
their presence is strongly felt by the charge distribution.

The minimum number of coefficients is easily found
when adequate values of the $\chi^2$
and of the confidence levels are reached.
It is more difficult to find the upper limit of the number of
expansion coefficients.
This problem  is rather well known in the
literature [3] and it has been
usually solved guessing the behavior of
the cross section, or of the form factor,
in the momentum transfer region beyond the maximum value
measured. The continuation of the cross
section in the unmeasured region fixes the values of
the high-order coefficients preventing them from producing
unwanted oscillation of the density distribution.

In spite of the more or less reasonable hypotheses done
about the behavior of the cross section in an unmeasured
region, it is clear that this procedure adds further
uncertainties to the charge distribution.

We believe that the charge density distributions should be
extracted considering only the measured data. The
truncation of the basis is intrinsically related to the
knowledge of the data on a limited range of momentum
transfer values, and this is part of the experimental
uncertainty.

The maximum number of coefficients determined by a set of
data extended up to $q_{max}$ is given in PWBA
by the eqs. (2.8b) and (2.20). These rules can be used as an
indication, but in the realistic description of the
scattering process they are not precise.

We have tackled the problem of cutting the expansion
together with the problem
of propagating the experimental errors of the cross section on
the charge distribution. The number of
expansion terms and the error on the charge distribution are both
fixed by the data.

The estimate of the
error on the density distribution is usually obtained
with the error propagation rules [3].
These rules assume a quadratic behavior of
the covariance matrix in the region of the values of
the parameters producing the minimum $\chi^2$.

We have calculated the error propagation with a different procedure.
{}From the original cross section calculated with the Woods-Saxon
densities, we have generated sets of pseudodata. Each set has been
constructed by random values normally distributed around
the exact cross section values with standard deviation equal to
the associated error.

Every fit produces a new charge distribution.
Putting
together all the charge distributions we obtain a band of lines
which automatically take into account the uncertainty
due to the errors of the cross sections.

Our procedure is equivalent to the standard procedure
if the quadratic
behavior of the covariance matrix around the point of
minimum of the fit is a good approximation.

As example, we show in figs. 5 and 6 some of the density
distributions bands obtained performing a set of 100 fits of
the randomly generated pseudodata of $^{40}Ca$.
We have applied this procedure for both FB and Hermite expansion.

Instead of drawing
every density with a continous line, we have indicated each
line with a fixed number of dots randomly distributed in
$r$. The darkening of one region gives
an indication of the fact that many charge distributions
are overlapping.

In these figures the full lines are representing the
Woods-Saxon densities used to generate the cross section
data.

We should point out that not all the lines which can be
drawn within the density band are reproducing a set of
pseudodata. The density points are strongly
correlated among them. For example, one of the constraints on these
points is the fact that the charge distribution
should satisfy the normalization condition (2.14).

Every fit produces also a new set of best fit parameters.
Superimposed to the figures of the distribution bands, for each
expansion coefficient, we show the absolute value of
the ratio between the standard deviation of
the n-th expansion coefficient
and its average value. When this ratio is above 1 the value of
the expansion coefficient is compatible with zero.
This means that the data are not sensitive to this term
of the expansion.

In each figure we have inserted the value of the average
$\chi^2$ per number of degrees of freedom
and the value of the average confidence
level.

In fig. 5 we observe that the fits done with $9$,$10$ and $11$ FB terms
have similar characteristics: the same value of the averaged $\chi^2$,
similar values of the averaged confidence levels and density bands
concentrated around the valued of the true distribution. The lower panel on
the right shows that the simple addition of another coefficient produces an
unstable situation. The density band is quite spread around the true value of
the density and $4$ out of $12$ coefficients are compatible with zero. This
instability of the fit it is not shown by the the values of the averaged
$\chi^2$ and confidence level.

This fact is not specific of the chosen expansion basis
because also in fig. 6, where the Hermite basis has been
used, we move suddenly from a stable situation to a
situation where 6 coefficients are compatible with zero.
We remark in fig. 6 that the presence of coefficients compatible with zero
does not mean that the density band is oscillating around the true
density distribution. Both the fits with $10$ and $11$ terms have one
term  compatible with zero, but they shows stable density bands. We
notice that the fit with 11 parameters produces an error band narrower
than the fit with 10. The fit with $13$ terms starts to show some
instability, but it is the fit with $14$ terms which produces a
strongly spread density band. In this last case $10$ out of $14$ terms
are undetermined. Note that, like in the FB case, the values of the
average $\chi^2$ and confidence level are not sensitive to the
instability of the fit.

A comparison between fig.5 and 6 emphasizes the differences between
the two bases. In fig. 5 we observe that the increase of the
number of coefficients of the FB expansion is affecting
only the last coefficients.
For example in the fit with 12 expansion terms, the last four
coefficients are undetermined, while the first eight have roughly the
same values presented in the other panels.
The situation is reversed
in the Hermite case where the addition of new coefficients strongly
modifies also the other ones, as we note on fig. 6.
This fact indicates that the Hermite coefficients are more strongly
correlated than the FB ones. This correlation is clearly indicated
by the covariance matrices obtained in the fits [14] and is
connected to the different characteristic of the two bases we have
already discussed in the description of fig.1. Each term of the FB
basis produces contributions rather well localized in momentum
space, while the contribution of each term of the Hermite basis is
spread over a wider momentum range.

The instability of the charge distributions is not connected to the fact
that high order terms are not constrained by the data, this is simply an
accident related to the FB expansion basis. There are general properties of
the charge distributions not restricted by the data, independently from the
chosen basis.

The tests performed for the $^{12}C$ and $^{208}Pb$ show
a clear jump in the values of the $\chi^2$ and of the
confidence level from a fit with a too small number of parameters to
a fit with an adequate number of parameters. This allows one to fix a
minimum number of expansion terms.

Unfortunately, as we have discussed for the $^{40}Ca$ case,
there is not such a clear
way to fix the maximum number of coefficients.
The criterion of excluding all the fits containing at least
one undetermined coefficient, i.e. compatible with zero, is
not adequate. As we have seen in the above
discussion, there are situations where some of the expansion
coefficients are compatible with zero but the distribution
band is concentrated around the Woods-Saxon distribution.

We can summarize our aim saying that
we are looking, in the space
of the coefficients, for a vector
which minimizes the $\chi^2$ and, at the same time, is stable
against statistical variations
of the cross section values.

The stability of a set of expansion coefficients
with respect to  the
variations of the cross section
can be measured by the quantity:
$$
S(N_c) = { { \sum_{n=1}^{N} < a_n >^2 -
           \sum_{n=1}^{N}  <a_n^2> }
 \over  {\sum_{n=1}^{N} <a_n>^2 }  } ,
\eqno(3.2)
$$
where $N$ is the number of expansion coefficients, and
$$
< a_n > = {1 \over M} \sum_{\mu=1}^{M}  a_n^{(\mu)}
\eqno(3.3)
$$
is the value of the coefficient $a_n$ averaged on $M$ samples, and
$$
< a^2_n > = {1 \over M} \sum_{\mu=1}^{M}  [a_n^{(\mu)}]^2 .
\eqno(3.4)
$$

A small value of the quantity $S(N)$ indicates that the vector
constructed with the average values of the coefficients is rather
well defined. This means that the errors on the coefficients are
quite small and this produces a narrow distribution band. Of course
large values of $S(N)$ are connected with broad distribution
bands.

Together with the quantity defined in eq. (3.2), which
gives information about
the stability of the fit with respect to statistical
variations, we should also include an indicator of the quality of
the fit.
For this reason  we define:
$$
G(N)= { <  cl(N) > \over S(N) } ,
\eqno(3.5)
$$
where $< cl(N) >$ is the average confidence level.
The quantity defined in eq. (3.5)
is not linked to the accuracy in the definition
of the single expansion coefficient.
For particular reasons one of the coefficients of the expansion
can be compatible with zero without
producing instabilities in the charge distribution.
The value of $G$ gives information about the global features of the
fit and the stability of the distribution density.

In fig. 7 we show the values of $G$, calculated for the
cases we have studied in figs. 5-6,
as a function of the number of expansion coefficients
$N$. The values of $G$ have been normalized in order to have all
the maxima equal to one.

The FB fits show sharp maxima, different for each
nucleus considered.
For the Hermite case the situation is more complicated.
The maxima are not so sharp and
there is a set of parameters with analogous
characteristics. This confirms the fact that the
Hermite expansion is more stable against variations of the
number of coefficients than the FB expansion.

The maxima of $G$ correspond to the ideal
number of expansion coefficients.  This number
optimizes the quality of the fit of the cross section data and the
stability of the result against the statistical
fluctuations.
In the next subsection we shall discuss the results we have obtained
applying this procedure
to the experimental data of ref. [10--13].

Before closing we would like to remark that the procedure we have
designed provides the best distribution band compatible with the
data, statistical errors included. However
this does not ensure that
the density generating the cross section certainly lies in
the distribution band obtained applying the procedure.

This point is quite well illustrated in fig.8. The full lines
represent a density generated with 15 FB coefficients.
We have used this density to produce $^{12}C$ pseudodata.
Since the shape of the density is quite odd, we verified that
the cross section was not showing anomalous behavior
at high $q$, at least
within a reasonable range of values of the momentum transfer.
For this reason
we have slightly increased the momentum transfer covered with
respect to the $^{12}C$ pseudodata used for the previous analysis.
With the charge distribution of fig.8
we generated data up a scattering angle of $140$ degrees
corresponding to a maximum momentum transfer of $3.6 fm^{-1}$.

We have applied the fit procedure to this set of
pseudodata.  We found an optimal value of
10 FB coefficients and this
produces the distribution band shown in the upper panel of fig. 8.
The original
charge distribution in not included in this band.
On the other hand it is shown in the central panel of fig. 8 that
the band generated by a 15 parameters fit also contains
the original distribution. The optimal fit done with 10 Hermite
functions shown in the bottom panel produces a distribution
band practically identical to that obtained with 10 FB
coefficients. This gives us confidence of the fact
that the procedure we have designed is
independent from the choice of the expansion basis and the result
is only related to the set of data.
This procedure fixes the number of expansion coefficients to obtain a
reasonable compromise between the quality of the fit and the
uncertainties on the density. This, however, does not ensure that the
real density is contained in the found band.

\vskip 1cm
{\sl 3.2 Application to experimental data}
\vskip 0.5cm

The methodology presented in the previous
section has been applied to the study of the experimental data
of $^{12}C$, $^{40}Ca$ and $^{208}Pb$
given
in refs. [10--13].

We have assumed that the quoted experimental
errors are only statistical and we
have generated sets of 100 angular distributions with the same
method used for the pseudodata. We considered
about the fact that we are
sampling on data which are already a statistical sample by
multiplying each error by the factor $\sqrt 2$. These data have
been fitted with
both the FB and Hermite basis changing the number of
expansion coefficients.

The results of these calculations are summarized in
tab. 3 where we present the average values of the $\chi^2$
and of the confidence level for each set of fits. We have performed
the  test of the $G$ factor defined in eq. (3.5) and we found,
for each expansion
basis, the optimal number of expansion term.

We present in figs. 9--11 the distribution bands produced by
the optimal number of coefficients for both expansion
bases and for the three nuclei considered. In these
figures the full lines are the charge distributions obtained with the
FB coefficients published in ref. [15]. For $^{12}C$ and
$^{40}Ca$ these densities present small differences with respect
to our density bands.

In fig. 11  together with our distribution
bands we present two charge distributions both obtained with FB
parametrizations quoted in ref. [15]. The full line is
produced by the 11 coefficients parametrization and it overlaps
completely with the distribution bands. The dashed line is instead
generated with the 17 coefficients parametrization.

We have performed a consistency test between the distribution bands
obtained with the FB and Hermite expansions and we found that, for each
nucleus considered, the differences between the FB and the Hermite
densities bands are compatible with zero. Looking in detail, however,
we found that they are not always
equally spaced around the zero, contrary to
the result we have obtained with the pseudodata.
This, together with the relatively low values of the confidence
levels shown in tab. 3,
can be an indication of the fact that the set of
data we have used are not a good statistical sample.

\vskip 1 cm

{\bf 4. Conclusions}
\vskip 0.5 cm

The work we have presented
in this paper has been addressed to the
study of the theoretical uncertainties in the
procedure of extracting the charge distributions from
elastic electron scattering cross sections. We
have worked within the direct scattering approach
to investigate the uncertainties of the the so-called
model independent techniques consisting in expanding
the charge distributions on a orthonormal basis and
finding the coefficients of the expansion to obtain
the best fit to the data.

In order
to obtain information independent from the choice of
the expansion basis we have used two bases with
different analytical properties. The FB basis is rather
localized in $q$ space, as it is shown in fig.1, and
it is spread in the coordinate space. The Hermite basis,
used for the first time in this context, has analogous
characteristics in both $q$ and $r$ space, the Fourier
transform of a Hermite function is still a Hermite
function. Fig.1 shows that the contribution of each
term of the Hermite expansion is not localized around a
specific value of $q$.

We found few advantages in using the Hermite basis
with respect to the FB ones. The Hermite expansion
seems to be more stable and usually it requires a
smaller number of expansion terms to obtain fits of the
same quality (see fig. 3 and 4). In addition, while with
FB one should impose that the distribution should be zero
after a certain radius, the Hermite expansion is not
requiring any hypothesis on the shape of the charge
density.

In any case the aim of our work was not the proposal of
a new expansion basis to be used in model independent
method, but rather the investigation of the
uncertainties related to the method in itself. We
found that in both the expansion bases it is not
possible to increase at will the number of expansion
terms. There is an upper limit when the uncertainty
band of the charge distribution starts to become very large and
shows big oscillations,
in spite of the fact that the quality of the fit to the
cross section  is not worsening.
The increase of the distribution uncertainty band appears when the
full set of expansion terms is not any more constrained by the data.

This problem is not related to the choice of the basis
but it is intrinsic to the extraction method, due to
the necessary discretization and limitation in $q$ of
the experimental data and to the presence of
statistical error. These are also the reasons why
the solution of the inverse problem is not
unique.

It is clearly possible to set up methods to
stabilize  the result of the fit.
For example, we have designed a procedure to determine
the number of coefficients giving the best compromise
between the quality of the fit and the stability of the
set of parameters with respect to statistical
fluctuations of the data. This procedure  consists in
repeating the fit with various numbers of expansion
terms. Each set of fits is analyzed with the indicator
defined in eq. (3.5) whose maximum provides  the
optimal number of coefficients.

This is only a useful pragmatical recipe but it does not
guarantee that the unknown quantity is within the
obtained density band. This is strikingly shown in
fig. 8 where the cross section produced by a
oscillating density distribution has been fitted. The
application of the prescription to both FB and Hermite
bases allows us to find an optimal value of the number
of expansion terms, but the  obtained
density bands are not containing the true density
distribution.

In the model independent method the truncation error
is unavoidable. It is even not possible to give a
realistic estimation of it, because it is related to
the procedure used to stabilize the density
distribution band.

The model independent method does not provide stable
solution of the inverse problem, in the sense that
small variations of the cross section produce large
variations of the density. It is necessary to perform an
opportune truncation of the basis to stabilize the
solution. The criteria used to perform the truncation
are based upon pragmatical considerations.
For these
reasons the distributions extracted
with these methods must be considered with caution when comparing to
nuclear models.

\vskip 1cm

It is a pleasure to acknowledge the useful discussions with
S.Caracciolo, A.M.Lallena, G.Lolos, G.Mancarella, R.Perrino, P.Rotelli,
R.Schiavilla and I.Sick.

\vfill\eject

{\bf References}
\vskip 1cm

\item{[1]} K.Chadan and P.C.Sabatier, Inverse Problems in Quantum
Scattering Theory, \par
 (Springer, Berlin, 1977) p.423. \par
\vskip .5cm

\item{[2]} J.L.Friar and J.W.Negele, Nucl. Phys. A212 (1973)
93.\par
\vskip .5cm

\item{[3]} B.Dreher, J.Friedrich, K.Merle, H.Rothhaas and
G.L\"urs,  Nucl. Phys. A235 (1974) 219.\par
\vskip .5cm

\item{[4]} J.Heisenberg and H.P.Blok, Ann. Rev. Nucl. Part. Sc.
33 (1983) 569.\par
\vskip .5cm

\item{[5]} D.W.L.Sprung, N.Yamanishi and D.C.Zheng, Nucl. Phys. A550
(1992) 89. \par
\vskip .5cm

\item{[6]} M.E.Grypeos, G.A.Lalazissis, S.E.Massen and C.P.Panos,
J.Phys. G 17 (1991) 1093. \par
\vskip .5cm

\item{[7]} R.E.Kozak, Am. J. Phys. 59 (1991) 74. \par
\vskip .5cm

\item{[8]} A.Messiah, Mecanique Quantique, vol 1
 (Dunod, Paris, 1962) p.418.
\vskip .5cm

\item{[9]} D.R.Yennie, D.G.Ravenhall and R.N.Wilson, Phys. Rev. 95
(1954) 500.\par
\vskip .5cm

\item{[10]} I.Sick and J.S.McCarthy, Nucl. Phys. A150 (1970) 631.\par
L.S.Cardman, J.W.Lightbody, S.Penner, S.P.Fivonzinsky
and X.K.Maruyama,\par
Phys. Lett. 91B (1980) 203.\par
W.Reuter, G.Fricke, K.Merle and H.Miska, Phys. Rev. C26 (1982)
806.\par
\vskip .5cm

\item {[11]} B.B.P.Sinha, G.A.Peterson, R.R.Whitney, I.Sick and
J.S.McCarthy,\par
Phys. Rev. C7 (1973) 1930.\par
I.Sick, J.Bellicard, J.M.Cavedon, B.Frois, M.Huet, P.Leconte,
P.X. Ho and \par
S.Platchkov,
 Phys. Lett. 88B (1979) 245.\par
\vskip .5cm

\item {[12]} J.Heisenberg, R.Hofstadter, J.S.McCarthy, I.Sick,
B.C.Clark, R.Herman and \par
D.G.Ravenhall, Phys. Rev. Lett. 23 (1969) 152.\par
B.Frois, J.Bellicard, J.M.Cavedon, M.Huet, P.Leconte, A.Nakada,
P.X.Ho and I.Sick,\par
Phys. Rev. Lett. 38 (1977) 1259.\par
\vskip .5cm

\item {[13]} J.M.Cavedon, Th\`ese de doctorat d'Etat, Paris 1980,
Unpublished.\par
\vskip .5cm

\item {[14]} P.Pellegrino, Tesi di laurea, Lecce 1994, Unpublished.\par
\vskip .5cm

\item{[15]} C.W.DeJager and C.DeVries, At. Data and Nucl. Data
Tables. 36 (1987) 495.

\vfill\eject


\def\tablerule{\noalign{\hrule}}
\def\qq{\;\;}
\midinsert{
$$\vcenter {\offinterlineskip \hrule
\def\spal1{height5pt
&\omit&&\omit&&\omit&&\omit&&\omit&\cr}
\halign { & \vrule# &
          $\qq\hfil#\hfil\qq$ & \vrule# &
          $\qq\hfil#\hfil\qq$ & \vrule# &
          $\qq\hfil#\hfil\qq$ & \vrule# &
          $\qq\hfil#\hfil\qq$ & \vrule# &
          $\qq\hfil#\hfil\qq$ & \vrule# &
          $\qq\hfil#\hfil\qq$ & \vrule# &
          $\qq\hfil#\hfil\qq$ & \vrule# &
          $\qq\hfil#\hfil\qq$ & \vrule# \cr
\spal1
& \omit && V_o \,\,\, (MeV) && V_{LS} \,\,\, (MeV) && R \,\,\, (fm) &&
a \,\,\, (fm) & \cr \spal1
\tablerule \spal1
& \omit && \omit && \omit && \omit &\cr \spal1
& ^{12}C  && \ 55.0
&& \ 3.2 && \ 2.86 &&  \ 0.57  &\cr \spal1
& \omit && \omit && \omit && \omit && \omit &\cr \spal1
\tablerule \spal1
& \omit && \omit && \omit && \omit && \omit &\cr \spal1
& ^{40}Ca  && \ 57.0 && \ 11.1 && \ 4.1 &&  \ 0.53  &\cr\spal1
& \omit && \omit && \omit && \omit && \omit &\cr
\spal1 \tablerule \spal1
& \omit && \omit && \omit && \omit &\cr \spal1
& ^{208}Pb  && \ 60.4 && \ 6.75 && \ 7.2 &&  \ 0.59  & \cr\spal1
& \omit && \omit && \omit && \omit &\cr \spal1
}\hrule}$$
\vskip7pt \noindent
\baselineskip=.15in
{\hsize 9cm{\bf Table~1.}\ \
Parameters of the Woods-Saxon potential (eq. 3.1) producing the
densities used to generate the pseudodata.
}}
\endinsert


\vskip 1cm

\def\tablerule{\noalign{\hrule}}
\def\qq{\;\;}
\midinsert{
$$\vcenter {\offinterlineskip \hrule
\def\spal1{height5pt
&\omit&&\omit&&\omit&&\omit&&\omit&\cr}
\halign { & \vrule# &
          $\qq\hfil#\hfil\qq$ & \vrule# &
          $\qq\hfil#\hfil\qq$ & \vrule# &
          $\qq\hfil#\hfil\qq$ & \vrule# &
          $\qq\hfil#\hfil\qq$ & \vrule# &
          $\qq\hfil#\hfil\qq$ & \vrule# &
          $\qq\hfil#\hfil\qq$ & \vrule# &
          $\qq\hfil#\hfil\qq$ & \vrule# &
          $\qq\hfil#\hfil\qq$ & \vrule# &
          $\qq\hfil#\hfil\qq$ & \vrule# \cr
\spal1
& \omit && n data && \epsilon \,\,\, (MeV) &&  \theta_{max} \,\,\, (deg.) &&
\ q_{max} \,\,\, (fm^{-1}) &\cr \spal1  \tablerule \spal1
& \omit && \omit && \omit && \omit && \omit &\cr \spal1
& ^{12}C  && \ 99
&& \ 400 && \ 132 &&  \ 3.5  &\cr \spal1
& \omit && \omit && \omit && \omit && \omit &\cr \spal1
\tablerule \spal1
& \omit && \omit && \omit && \omit && \omit &\cr \spal1
& ^{40}Ca  && \ 148
&& \ 502 && \ 110 &&  \ 3.9  &\cr
\spal1
& \omit && \omit && \omit && \omit && \omit &\cr
\spal1 \tablerule \spal1
& \omit && \omit && \omit && \omit && \omit &\cr \spal1
& ^{208}Pb  && \ 148
&& \ 502 && \ 87 &&  \ 3.2  &\cr
\spal1
& \omit && \omit && \omit && \omit && \omit &\cr \spal1
}\hrule}$$
\vskip7pt \noindent
\baselineskip=.15in
{\hsize 9cm{\bf Table~2.}\ \
Information about the pseudodata. For each nucleus considered we
present: the number of pseudodata generated ($ndata$),  the
energy of the electrons  ($\epsilon$), the maximum scattering
angle ($\theta_{max}$) and  the corresponding maximum
momentum transfer  ($q_{max}$).
 }}
\endinsert

\vfill\eject


\vskip 1cm

\def\tablerule{\noalign{\hrule}}
\def\qq{\;\;}
\midinsert{
$$\vcenter {\offinterlineskip \hrule
\def\spal1{height5pt
&\omit&&\omit&&\omit&&\omit&&\omit&&\omit&&\omit&&\omit&&\omit&
&\omit&&\omit&&\omit&&\omit&\cr}
\halign { & \vrule# &
          $\qq\hfil#\hfil\qq$ & \vrule# &
          $\qq\hfil#\hfil\qq$ & \vrule# &
          $\qq\hfil#\hfil\qq$ & \vrule# &
          $\qq\hfil#\hfil\qq$ & \vrule# &
          $\qq\hfil#\hfil\qq$ & \vrule# &
          $\qq\hfil#\hfil\qq$ & \vrule# &
          $\qq\hfil#\hfil\qq$ & \vrule# &
          $\qq\hfil#\hfil\qq$ & \vrule# &
          $\qq\hfil#\hfil\qq$ & \vrule# &
          $\qq\hfil#\hfil\qq$ & \vrule# &
          $\qq\hfil#\hfil\qq$ & \vrule# &
          $\qq\hfil#\hfil\qq$ & \vrule# &
          $\qq\hfil#\hfil\qq$ & \vrule# \cr

&\cr   & \omit&height5pt & \multispan7 && \multispan7 && \multispan7 &
 \cr
& \omit && \multispan7\hfil  $^{12}C$ \hfil && \multispan7\hfil $^{40}Ca$ \hfil
&& \multispan7\hfil $^{208}Pb$ \hfil
&\cr   & \omit&height5pt & \multispan7 && \multispan7 && \multispan7 &
\cr
\tablerule

&\cr   & \omit&height5pt & \multispan3 && \multispan3 && \multispan3 &&
\multispan3
&& \multispan3 && \multispan3 &
 \cr
& \omit && \multispan3\hfil FB \hfil && \multispan3\hfil Hermite \hfil
&& \multispan3\hfil FB \hfil && \multispan3\hfil Hermite \hfil
&& \multispan3\hfil FB \hfil && \multispan3\hfil Hermite \hfil
&\cr   & \omit&height5pt & \multispan3 && \multispan3 && \multispan3 &&
\multispan3 &
& \multispan3 && \multispan3 &
\cr
\tablerule

\cr
\spal1
& \ N && \chi^2 && \ c.l.  &&  \chi^2 && \ c.l. && \chi^2
&& \ c.l. && \chi^2  && \ c.l. && \chi^2 && \ c.l. && \chi^2 && \ c.l. & \cr
\spal1
\tablerule \spal1

& \omit && \omit && \omit && \omit && \omit && \omit && \omit && \omit &
& \omit &&\omit&&\omit&&\omit&&\omit&\cr \spal1
& \ 7  && \omit
&& \omit && \ 1.08 &&  \ 0.35  && \omit &&  \omit && \ 1.17 && \ 0.17 &
& \omit && \omit && \omit && \omit &\cr \spal1
& \omit && \omit && \omit && \omit && \omit && \omit && \omit && \omit
&& \omit &&\omit&&\omit&&\omit&&\omit&\cr \spal1

& \omit && \omit && \omit && \omit && \omit && \omit && \omit && \omit && \omit
&
&\omit&&\omit&&\omit&&\omit&\cr \spal1
& \ 8  && \ 1.93
&& \ 0.00 && \ 1.09 &&  \ 0.35  && \ 1.33 &&  \ 0.03 && \ 1.15 && \ 0.19 &
& \omit  &&\omit&& \omit && \omit &\cr
\spal1
& \omit && \omit && \omit && \omit && \omit && \omit && \omit && \omit &
&\omit&&\omit&&\omit&& \omit &&\omit&\cr \spal1

& \omit && \omit && \omit && \omit && \omit && \omit && \omit &
&\omit&&\omit&&\omit&& \omit &&\omit&\cr \spal1
& \ 9  && \ 1.18
&& \ 0.19 && \ 1.07 &&  \ 0.39  && \ 1.12 &&  \ 0.26 && \ 1.12 && \ 0.26 &
& \ 3.31 && \ 0.00 && \ 1.47 && \ 0.00 &\cr
\spal1
& \omit && \omit && \omit && \omit && \omit && \omit && \omit && \omit && \omit
&
&\omit&&\omit&&\omit&&\omit&\cr \spal1

& \omit && \omit && \omit && \omit && \omit && \omit && \omit &
&\omit&&\omit&&\omit&& \omit &&\omit&\cr \spal1
& \ 10  && \ 1.13
&& \ 0.27 && \ 1.08 &&  \ 0.37  && \ 1.13 &&  \ 0.25 && \ 1.12 && \ 0.27 &
& \ 1.17 && \ 0.17 && \ 1.33 && \ 0.02 &\cr
\spal1
& \omit && \omit && \omit && \omit && \omit && \omit && \omit && \omit && \omit
&
&\omit&&\omit&&\omit&&\omit&\cr \spal1

& \omit && \omit && \omit && \omit && \omit && \omit && \omit &
&\omit&&\omit&&\omit&& \omit &&\omit&\cr \spal1
& \ 11  && \ 1.13
&& \ 0.28 && \ 1.10 &&  \ 0.33  && \ 1.11 &&  \ 0.15 && \ 1.13 && \ 0.24 &
& \ 1.17 && \ 0.19 && \ 1.07 && \ 0.24 &\cr
\spal1
& \omit && \omit && \omit && \omit && \omit && \omit && \omit && \omit && \omit
&
&\omit&&\omit&&\omit&&\omit&\cr \spal1

& \omit && \omit && \omit && \omit && \omit && \omit && \omit &
&\omit&&\omit&&\omit&& \omit &&\omit&\cr \spal1
& \ 12  && \ 1.14
&& \ 0.27 && \ 1.08 &&  \ 0.34  && \omit &&  \omit && \omit && \omit &
& \ 1.14 && \ 0.22 && \ 1.07 && \ 0.24 &\cr
\spal1
& \omit && \omit && \omit && \omit && \omit && \omit && \omit && \omit && \omit
&
&\omit&&\omit&&\omit&&\omit&\cr \spal1

& \omit && \omit && \omit && \omit && \omit && \omit && \omit &
&\omit&&\omit&&\omit&& \omit &&\omit&\cr \spal1
& \ 13  && \omit
&& \omit && \omit &&  \omit  && \omit &&  \omit && \omit && \omit &
& \ 1.16 && \ 0.20 && \ 1.10 && \ 0.19 &\cr
\spal1
& \omit && \omit && \omit && \omit && \omit && \omit && \omit && \omit && \omit
&
&\omit&&\omit&&\omit&&\omit&\cr \spal1

& \omit && \omit && \omit && \omit && \omit && \omit && \omit &
&\omit&&\omit&&\omit&& \omit &&\omit&\cr \spal1
& \ 14  && \omit
&& \omit && \omit &&  \omit  && \omit &&  \omit && \omit && \omit &
& \ 1.13 && \ 0.23 && \ 1.08 && \ 0.22 &\cr
\spal1
& \omit && \omit && \omit && \omit && \omit && \omit && \omit && \omit && \omit
&
&\omit&&\omit&&\omit&&\omit&\cr \spal1
}\hrule}$$
\vskip7pt \noindent
\baselineskip=.15in
{\hsize 9cm{\bf Table~3.}\ \
Averaged $\chi^2$ and confidence level for the fits to the experimental data
obtained
with various number of expansion coefficients.
 }}
\endinsert

\vfill\eject

{\bf Figure captions}

\vskip .5cm

\item {Fig.1}
Form factor squared shown as a function of the
momentum transfer.
The thick lines represent the form factor produced
by the test density distribution of eq. (2.11).
The panels $a$ and $c$ show the form factor obtained with
$4$ and $6$ FB expansions terms respectively. The panels $b$ and $d$
show the analogous fit performed with the Hermite expansion basis.
The full thin lines show the form factors generated summing the N
terms of the expansion and the dashed line show the contribution of
the n-th term. The form factors are all normalized such as
$F(0)=1.0$ .
\vskip .2cm

\item {Fig.2}
Comparison between exact form factors, eq. (2.12), (thick lines)
and form factors calculated with a truncated density, eq.
(2.15), (thin lines).
\vskip .2cm

\item {Fig.3}
Charge densities obtained from a FB fit of the $^{12}C$ pseudodata done with
various number of expansion terms. The results of the upper panel
have been obtained with $R_c=8$ fm, those of the lower panel with $R_c=10$ fm.
The full lines represent the Woods-Saxon charge distribution
used to generate the pseudodata.
\vskip .2cm

\item {Fig.4}
The same as fig. 3 for the Hermite expansion basis. Also in this
case we presents results obtained with two different values
of the scale parameters: $\beta=1.5$ fm (upper panel) and
$\beta=1.7$ fm (lower panel).
\vskip .2cm

\item {Fig.5}
The bottom and left scale refer to the
distribution bands produced by the 100 FB fits to $^{40}Ca$
pseudodata represented by the randomly scattered dots ($R_c=8.0$
fm). The full lines show the Woods-Saxon charge distribution
generating the pseudodata. The upper scale shows the
number of expansion coefficients used in the fit and the right
scale refers to the absolute value of the ratio between the
error and the average value of the expansion coefficient. These
ratios are shown by the circles. The average values of the
reduced $\chi^2$ and of the confidence level are also given.
\vskip .2cm

\item {Fig.6}
The same as fig. 5 for the case of the Hermite expansion
($\beta=1.5$ fm).
\vskip .2cm

\item {Fig.7}
The quantity $G$ defined in eq. (3.5) as a function of the
expansion coefficients for the three nuclei under investigation. The
full lines are related to $^{12}C$, the dotted lines to $^{40}Ca$
and the dashed lines to $^{208}Pb$. Each line has been normalized in
order to have the maximum equal to 1.
\vskip .2cm

\item{Fig.8}
Fit to the pseudodata generated by the charge distribution shown
by the full lines. The distribution band of the upper panel has
been obtained with 10 FB coefficients and the density band of
the middle panel with 15 FB coefficients. The lower panel shows
a density band obtained with 10 Hermite coefficients.
\vskip .2cm

\item {Fig.9}
Charge density bands obtained from the fit to the experimental data of
ref.[10] using the FB basis (10 coefficients) and the Hermite basis (10
coefficients). The full lines represent the density obtained by
the FB parametrization given in ref.[15].
\vskip .2cm

\item {Fig.10}
Same as fig. 9 for $^{40}Ca$. The experimental data are from
refs. [11] and we used 10  coefficients for both FB and Hermite
basis. Like in fig. 9 the full lines are giving the FB density of
ref.[15].
\vskip .2cm

\item {Fig.11}
Same as the previous two figures for $^{208}Pb$ where we used
the data of refs. [12,13]. The number of coefficients used
in the fits are 12 and 11 for the FB and the Hermite basis
respectively.
Full and dashed lines are showing the distributions given
by the FB parametrization in ref. [15]. The full  lines
have been obtained with the 11 FB coefficients
parametrization, the dashed lines with the 17 FB coefficients
parametrization.
\vskip .2cm

\vfill\eject

\end